\documentclass[proof]{WileyASNA-v1}

\articletype{Article Type}%

\received{26 April 2016}
\revised{6 June 2016}
\accepted{6 June 2016}

\begin{document}

\title{Updated Absolute Parameters and Kinematics of IS CMa}

\author[1]{S. Evcil*}

\author[2,3]{S. Adalalı}

\author[4]{N. Alan}

\author[1]{R. Canbay}

\author[4]{S. Bilir}

\authormark{Evcil \textsc{et al.}}

\address[1]{\orgdiv{Institute of Graduate Studies in Science}, \orgname{Istanbul University}, \orgaddress{\state{Istanbul}, \country{Türkiye}}}

\address[2]{\orgdiv{School of Graduate Studies}, \orgname{Çanakkale Onsekiz Mart University}, \orgaddress{\state{Çanakkale}, \country{Türkiye}}}

\address[3]{\orgdiv{Astrophysics Research Center and Ulupınar Observatory}, \orgname{Çanakkale Onsekiz Mart University}, \orgaddress{\state{Çanakkale}, \country{Türkiye}}}

\address[4]{\orgdiv{Faculty of Science, Department of Astronomy and Space Sciences}, \orgname{Istanbul University}, \orgaddress{\state{Istanbul}, \country{Türkiye}}}

\corres{*Selçuk Bilir, Faculty of Science, Department of Astronomy and Space Sciences, Istanbul, Türkiye. \email{sbilir@istanbul.edu.tr}}

%\presentaddress{This is sample for present address text this is sample for present address text}

\abstract{Eclipsing binary systems are significant objects for astrophysics in that direct observations can determine the fundamental parameters of stars. In this study, we determined precisely the fundamental parameters of the binary component stars obtained by simultaneous analysis of radial velocities and the {\it TESS} light curve using the Wilson and Devinney code. Following the analysis, the masses and radii of the primary and secondary components were determined as $M_{1}= 1.58\pm 0.01M_\odot$, $M_{2}= 0.48\pm0.02M_\odot$, and $R_{1}=1.93\pm 0.01R_\odot$, $R_{2}= 1.14\pm 0.01 R_\odot$, respectively. Furthermore, the distance of IS CMa is calculated as $92.7\pm6.5$ pc. On the basis of the analysis of the mid-eclipse times, it was found that the variation in the orbital period is represented by an upward parabola. It has an increasing rate of $dP/dt$ = 1.09 $\times$ 10$^{-7}$ day yr$^{-1}$. Using PARSEC stellar evolutionary tracks and isochrones with solar metallicity were estimated the age of IS CMa as $1.3\pm0.1$ Gyr. Kinematic and Galactic orbital parameters of IS CMa were obtained from the astrometric and spectroscopic data of the system. The Galactic orbit analysis reveals that IS CMa formed inside the solar circle and it is a member of the young thin-disc population.}


%\jnlcitation{\cname{%
%\author{Williams K.}, 
%\author{B. Hoskins}, 
%\author{R. Lee}, 
%\author{G. Masato}, and 
%\author{T. Woollings}} (\cyear{2016}), 
%\ctitle{A regime analysis of Atlantic winter jet variability applied to evaluate HadGEM3-GC2}, \cjournal{Q.J.R. Meteorol. Soc.}, \cvol{2017;00:1--6}.}

\fundingInfo{Türkiye Bilimsel ve Teknolojik Araştırma Kurumu, Grant/Award Number:119F072}

\maketitle

%\footnotetext{\textbf{Abbreviations:} ANA, anti-nuclear antibodies; APC, antigen-presenting cells; IRF, interferon regulatory factor}

\section{Introduction}\label{sec1}

\label{introduction}
In the determination of fundamental stellar parameters (effective temperature $T_{\rm eff}$, mass $M$, radius $R$, luminosity $L$), eclipsing close binary stars are privileged because of their observational convenience \citep{Eker2024}. These are systems consisting of at least two stars bound together by gravitational forces and orbiting around a common center of mass according to Kepler's laws. Their physical parameters can be obtained directly by analysing their light and radial velocity curves over time under certain theories \citep{Kreiner2003, Zola2010}. To obtain accurate physical parameters, high-resolution spectroscopic and high-precision photometric observations as well as fast, reliable, and realistic model-based solution methods should be used \citep{Eker2015, Eker2018, Eker2021}. 

The evolution, mass loss, and interactions of close binary systems can be investigated by analysing observational data of contact binary systems. The periods of these systems are usually between $0.2$ and 1 day, and the mass ratios of their components are usually between $0.1$ and $0.5$ \citep{Yakut2005, Bilir2005, Eker2008}. The surface temperatures of the component stars are generally the same due to the common convective envelope formation. This situation leads us to expect the same mass values for both component stars. However, the situation is quite different, especially in the W UMa systems, which are called low-temperature contact binaries (LTCBs). The components of W UMa are usually located in the main sequence or below the zero age main sequence (ZAMS). However, this presented an important difficulty in understanding their fundamental physical characteristics. Since the effective temperature of main-sequence stars is directly correlated with their mass, the temperatures of components with various masses are also different \citep{Soydugan2015}. The first classification of the contact binary systems was made by \citet{Kuiper1941}. Later, \citet{Kopal1955} classified them according to whether Roche lobes are filled or not. The classification of W UMa binaries into A and W types was made by \citet{Lucy1968} and \citet{Binnendijk1970}. Type A W UMas have relatively more massive component stars, and the primary component is more massive and hotter. The component stars of W-type W UMas are comparatively low-mass, and the primary component is lower mass and hotter. Eclipsing close binary systems allows us to test fundamental theories and understand stellar evolution and stellar structure \citep{Yuan2019}.

The formation of contact binaries is complex and involves mechanisms like magnetic breaking, tidal locking, and dissipation of orbital angular momentum. Contact binaries can be classified into different subtypes based on their characteristics, such as orbital inclination and photometric parameters. The energy transfer in W UMa-type stars can have a substantial impact on their evolution, leading to varied results depending on the location of the energy transfer within the shared envelope. A-type systems exhibit longer orbital periods, higher temperatures, and a lower mass ratio in comparison to W-type systems \citep{Rucinski1983, Smith1984}. Contact binaries characterised by a low total mass tend to undergo a transformation towards smaller mass ratios and more profound shared envelopes. This envelope is formed when one of the binary components overflows its Roche lobe, causing the mass to be transferred to the other star. W UMa type close binaries consist of secondary components that feature unique characteristics or track different paths of evolution in comparison to the primary components. The secondary components exhibit higher luminosity and larger size compared to their main-sequence neighbors. There remains a discussion on whether the subtypes can be considered an evolutionary sequence based on their effective temperatures \citep{Papa1979, Yildiz2013}. The topic of discussion revolves around the transfer of energy from primary stars to secondary stars. The energy transfer area has a significant impact on the structure and development of contact binaries, exerting distinct effects on A and W-type systems. Systems characterised by low total mass and belonging to the A-type classification can be regarded as more evolutionary stages of W-subtypes. The secondary star's temperature may exceed that of the expected values due to factors such as expansion or compression \citep{Li2004, Jiang2009, Zhang2020, Qian2020}.
\par
IS Canis Major (HD 44524, HIP 30174, SAO 171609, TIC 49547177; $l=237^{\rm o}.184639,~~b=-19^{\rm o}.166833$), an eclipsing binary star of the W UMa type, exhibits considerable brightness and is characterized by a short orbital period of 0.616982 days. Observable only from the Southern Hemisphere, the system's light variation was first detected by the {\it Hipparcos} satellite \citep{ESA1997}. \citet{Adelman2001} compiled a list of 2027 stars with the largest photometric amplitudes based on {\it Hipparcos} photometry, containing IS CMa. Several catalogues, such as those by \citet{Olsen1994}, \citet{Carrasco1995}, and \citet{Malkov2006}, provide photometric properties for IS CMa. \citet{Rucinski2002} included IS CMa in a sample of bright contact binary stars up to a limit of 7.5 mag, emphasizing the need for further investigation. \citet{Selam2004} determined the geometric elements (mass ratio ($q$), degree of contact ($f$), and inclination ($i$)) of selected systems using {\it Hipparcos} photometry, providing values of $q=0.30$, $f=0.50$, and $i=75^{\rm o}$ for IS CMa. \citet{Nordstrom2004} presented comprehensive data on metallicity, rotation, age, kinematics, and Galactic orbits for a sample of 16,682 F and G dwarf stars, citing a logarithmic [Fe/H] of -0.36 dex, a distance of 100 pc, and an age of 1.7 Gyr for IS CMa. \citet{Rucinski2006} attempted a period-luminosity calibration for 21 contact binaries based on good {\it Hipparcos} parallaxes but excluded IS CMa due to its outlier status. \citet{Pribulla2006} listed IS CMa among 151 contact systems with potential additional components, urging further investigations into these candidates for multiplicity. \citet{Ozkardes2010} obtained the first high-resolution spectroscopic observations of the IS CMa and determined its orbital parameters of the system. The researchers performed the light and radial velocity curves solution of the system and determined the fundamental parameters of the components as $M_{1}= 1.68\pm 0.04\, M_\odot$, $M_{2}= 0.50\pm0.02\, M_\odot$, $R_{1}=2.00\pm 0.02\,R_\odot$, $R_{2}= 1.18\pm 0.03\, R_\odot$, and $L_{1}=7.65\pm 0.60\, R_\odot$, $L_{2}= 1.99\pm 0.80\, R_\odot$, respectively.

With the introduction of space-based telescopes for photometric, astrometric, and spectroscopic observations, much more precise data is being obtained. This allows for more accurate and reliable determination of the fundamental stellar parameters based on observational findings than ground-based observations. In particular, the Transiting Exoplanet Survey Satellite ({\it TESS}) scanned the all-sky survey and obtained highly precise light curves with eclipsing binaries, pulsating stars, and exoplanets has been a breakthrough in stellar astrophysics \citep{Ricker2015}. Moreover, the Global Astrometric Interferometer for Astrophysics ({\it Gaia}) satellite's measurement of precise astrometric data, including faint stars, allows for the measurement of distance, which is a fundamental problem of astronomy in other words, the precise determination of the luminosity of stars \citep{Gaia2016}. The precise measurements obtained by these two space telescopes have led to a re-investigation of the IS CMa system.

In this study, photometric and astrometric observations from the {\it TESS} and {\it Gaia} space telescopes are combined with ground-based spectroscopic data to derive the fundamental stellar parameters of the IS CMa. Furthermore, the kinematic and dynamical orbital parameters of this system were calculated to determine the Galactic population type and birthplace of IS CMa.

\section{Observational Data}
\label{observational data}
The analysis of the light curve involved harnessing data obtained from the {\it TESS}. {\it TESS} systematically surveys vast portions of the celestial sphere through segmented sectors, dedicating a focused observation period of 27.4 days per sector. Operating within 600-1000 nm wavelength range, {\it TESS} delivers comprehensive broadband photometric data, as detailed by \citet{Ricker2015}.

The specific {\it TESS} data pertains to IS CMa and originates from Sector 6 (2018-12-12 and 2019-01-06) and Sector 33 (2020-12-17 and 2021-01-13) with each exposure time 120 seconds. These observational records were obtained from the Mikulski Archive for Space Telescopes (MAST)\footnote{https://archive.stsci.edu/} database. During our photometric analysis, we utilized Pre-search Data Conditioning Simple Aperture Photometry (PDCSAP) light curves, a methodology introduced by \citet{Ricker2015}. The application of PDCSAP facilitated the refinement and preparation of the photometric data for in-depth examination. Notably, the mean error associated with the photometric measurements was approximately 0.1\%. This meticulous approach ensures the reliability and precision of the derived light curve information, contributing to the robustness of our findings in the domain of astrophysical investigations.

The radial velocity data of the IS CMa were taken from \citet{Ozkardes2010}. They measured the radial velocity values by using the spectra with $R=40,000$ resolution in the analysis.In \citet{Ozkardes2010} study, there are 15 and 12 radial velocity measurements of IS CMa's primary and secondary stars, respectively. 

\section{Orbital Period Variations}
\label{period}
Orbital period change investigations have provided valuable insights into the dynamics and evolution of the contact binary star systems \citep{Qian2020, Zhang2020}. They have shown that the mass transfer between the two stars in contact binaries plays a crucial role in shaping their evolution and can lead to phenomena such as common envelope phases and mass loss. Orbital period changes may occur when a binary's orbital period decreases or increases due to various factors such as mass transfer, tidal interactions, and angular momentum exchange. Understanding the mechanisms behind period changes in contact binaries is essential for studying their long-term evolution and predicting their fate \citep{Zhou2016, Li2021, Liao2019}. Hence, the orbital period variations of the contact binary system IS CMa were examined for the first time. We have thoroughly examined all the available times in the literature and have calculated new minimum times based on data from the {\it TESS} satellite (Table \ref{tab:min_times}). A total of 21 minimum times were gathered, with 9 minima previously reported in the {\it O-C} gateway\footnote{http://var2.astro.cz/EN/} \citep{Pas2006} and we have obtained 12 new minima from the TESS satellite. The {\it TESS} observations include Sector 6 and Sector 33, capturing data with a 120-second exposure duration and covering the period from 2018 to 2020. The {\it TESS} minima and the related errors were calculated using the \citet{Kwee1956} technique and presented in Table \ref{tab:min_times}.

% Table 1
\begin{table}
\setlength{\tabcolsep}{1.2pt}
%\centering
\captionsetup{justification=centering}
\caption{{\it O-C} data for {IS CMa}. The mid-eclipse timing, eclipse types (p: primary, s: secondary), calculated epoch, observed minus the computed and its residuals in Equation \ref{eq:O-C}.}
    \begin{tabular}{lccccccc}
\hline
\text{JD(Hel.)} & \text{Type} & \text{Epoch} & \text{{\it O-C}} & \text{Residual} &\text{Reference}\\
(2400000+)0 &  &  & (day) & (day) &\\
\hline
48500.35900 &\text{p}  & -6483  & 0.00238 & -0.00100 &\text{{\it Hipparcos}}$^1$\\ 
52787.47000 &\text{s}  & 465.5  & -0.00001& -0.00002 &\text{Paschke A.}$^1$\\
54042.72490 &\text{p}  & 2500   & 0.00093 &  0.00042 &\text{Ogloza W.}$^1$\\
54054.75790 &\text{s}  & 2519.5 & 0.00274 &  0.00222 &\text{Ogloza W.}$^1$\\
54071.72390 &\text{p}  & 2547   & 0.00168 &  0.00115 &\text{Ogloza W.}$^1$\\
56621.72300 &\text{p}  & 6680   & 0.00587 &  0.00227 &\text{Paschke A.}$^1$\\
57817.44000 &\text{p}  & 8618   & 0.00786 &  0.00188 &\text{Paschke A.}$^1$\\
58468.35621 &\text{p}  & 9673   & 0.00594 & -0.00159 &\text{{\it TESS}}$^2$\\
58468.66508 &\text{s}  & 9673.5 & 0.00632 & -0.00119 &\text{{\it TESS}}$^2$\\
58476.06891 &\text{s}  & 9685.5 & 0.00634 & -0.00121 &\text{{\it TESS}}$^2$\\
58476.37712 &\text{p}  & 9686   & 0.00606 & -0.00151 &\text{{\it TESS}}$^2$\\
58489.64136 &\text{s}  & 9707.5 & 0.00514 & -0.00240 &\text{{\it TESS}}$^2$\\
58489.95061 &\text{p}  & 9708   & 0.00590 & -0.00169 &\text{{\it TESS}}$^2$\\
59201.95341 &\text{p}  & 10862  & 0.00915 & -0.00035 &\text{{\it TESS}}$^2$\\
59202.26196 &\text{s}  & 10862.5& 0.00921 & -0.00024 &\text{{\it TESS}}$^2$\\
59207.81451 &\text{s}  & 10871.5& 0.00891 & -0.00061 &\text{{\it TESS}}$^2$\\
59208.12311 &\text{p}  & 10872  & 0.00901 & -0.00051 &\text{{\it TESS}}$^2$\\
59213.36754 &\text{s}  & 10880.5& 0.00908 & -0.00049 &\text{{\it TESS}}$^2$\\
59213.67594 &\text{p}  & 10881  & 0.00899 & -0.00058 &\text{{\it TESS}}$^2$\\
59655.44600 &\text{p}  & 11597  & 0.01850 &  0.00768 &\text{Paschke A.}$^1$\\
60019.46400 &\text{p}  & 12187  & 0.01593 &  0.00398 &\text{Paschke A.}$^1$\\
\hline
  \end{tabular}
\\
Ref: [1] {\it O - C} Gateway, \url{http://var2.astro.cz/ocgate/}, [2] This study.
    \label{tab:min_times}
\end{table} 
%\par
The {\it O-C} investigations represent that linear and quadratic (parabola fitting) assumptions, are often employed to study period variations in contact binaries. Period occulations can be identify anomalies and trends in the {\it O-C} diagrams over time. We utilized the {\it O-C} approach frequently used by researchers in our study to investigate possible variations in the IS CMa system \citep{Soydugan2008, Soydugan2013, Han2024}. So we used Equation \ref{eq:HJD1}, where $T_0$ is the primary minima time, $P_0$ is the orbital period, and {\it E} is the number of cycles, to {\it O-C} variation of IS CMa.
% Eq1
\begin{equation}
\label{eq:HJD1}
\text{MinI}\text{(HJD)}=T_0+P_0\times E.   
\end{equation}
%\par
Therefore, we used the initial linear ephemerises which are $T_0$ = 2452500.2633 and {\it E} = 0.61698502 from \citet{Kreiner2004}. Overall, we used all the minimum times, which are distributed over about 32 years. As part of the data analysis, the differences between the observed and calculated {\it O-C} values were examined. The 21 minimum times were mostly gathered from CCD data. The data included only of CCD observations and were analysed using a {\sc MATLAB} software provided by \cite{Zasche2009}. The weights assigned to individual observations for CCD data were established at 10, considering the errors, as advised by \cite{Zasche2009}. Utilizing this data, we made adjustments to the ephemeris and {\it O-C} values obtained according to the Equation \ref{eq:O-C}.
% Eq2
\begin{equation}
\begin{split}
\label{eq:O-C}
\text{MinI}\text{(HJD)}&=2452500.26433(6)+0.6169838471(1)\times E\\
&+0.92(5)\times10^{-10}\times E^2.\\ 
\end{split}
\end{equation}
\par
\begin{figure}
\centering\includegraphics[width=1\linewidth]{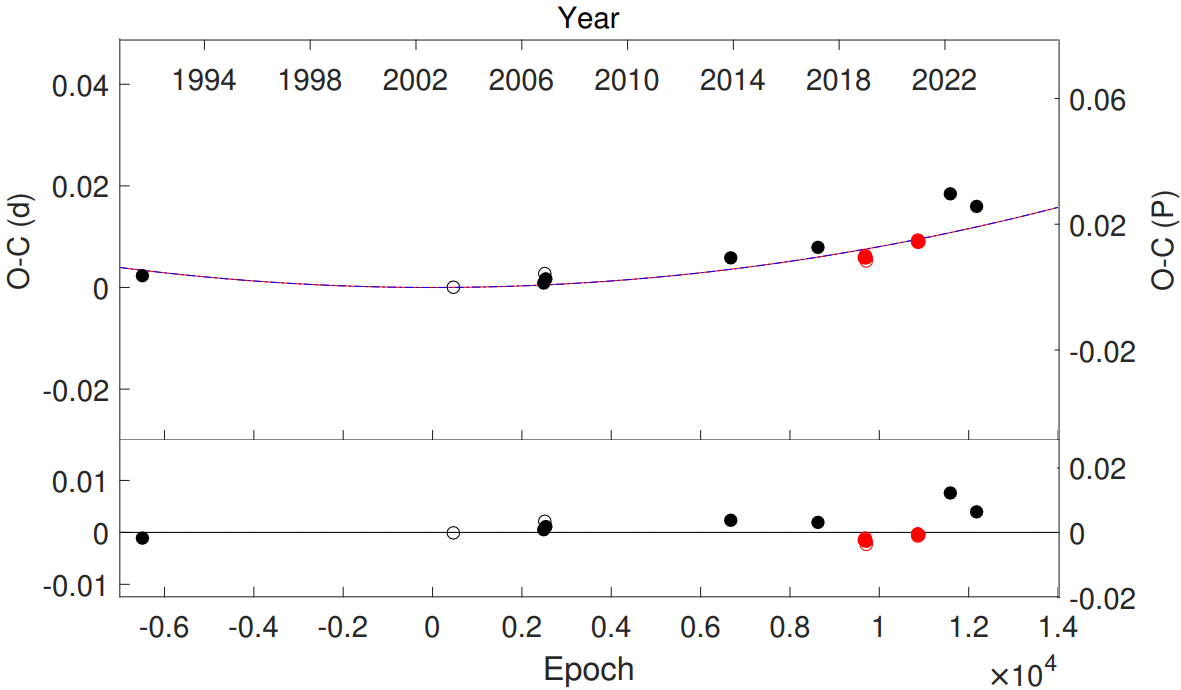}
\caption{The {\it O-C} plot for IS CMa displays the quadratic fit as a continuous line in the upper panel. This analysis utilizes literature-derived times and the red minimum times obtained from the {\it TESS} satellite. The corresponding residuals are plotted in the lower panel.}
    \label{fig:O-C}
\end{figure}

We generated the IS CMa {\it O-C} diagram using this equation, as seen in Figure \ref{fig:O-C}. We have gathered and examined all the times of minima for IS CMa so our investigation of the {\it O-C} curve, a noticeable parabolic variation is seen in the orbital period. The final values are given in Table \ref{tab:o-c_parameters} and the {\it O-C} values with residuals are listed in Table \ref{tab:min_times}. The final residuals also demonstrated in the lower panel of Figure \ref{fig:O-C}. {\it O-C} versus the time of observation (epoch) exhibits an upward parabolic structure.
The observed variations in orbital period in contact binaries are caused by mass transfer from the less massive component to the more massive one, resulting in an increase in the period \citep{Qian2001}. We have determined the quadratic coefficient $Q$ (day) = 0.92(5) $\times$ 10$^{-10}$ in Equation \ref{eq:O-C}, so for the long-term increasing period rate was calculated around $dP/dt=1.09(8)\times10^{-7}$ days per year. Mass transfer occurs as a result of the binary system's contact configuration, and the positive value of the quadratic term indicates a conservative transfer of mass from the secondary to the primary component. The increase in the period variation is attributed to mass transfer from the less massive component to the more massive component. The transfer rate can be determined using the \citet{Hilditch2001} equation,
% Eq3_1
\begin{equation}
\begin{split}
\label{eq:mass}
\dot{M_1}&=\frac{\dot{P}}{3P}\frac{M_1M_2}{(M_1-M_2)},\\ 
\end{split}
\end{equation}
so we determined a mass transfer rate $dM/dt=0.40(5)\times10^{-7}$ $M_{\odot}$ per year for the IS CMa contact binary. The parameters used are the exact ones derived from this investigation, which provide accurate and reliable insights into IS CMa. This indicates that, in addition to the secular increase in the orbital period, there may also be a possible periodic variation \citep{Qian2001, Gazeas2008, Liu2018}.

% Figure 1

% Table 2
\begin{table}[]
\centering
\caption{Results derived from the {\it O-C} analysis.}
    \begin{tabularx}{\columnwidth} { 
   >{\raggedright\arraybackslash}X  >{\raggedleft\arraybackslash}X}
     \toprule
\text{Parameter}                 &\text{Value}\\
\midrule
$T_0$ (HJD-2400 000)             &52500.26433 (6)\\
$P_{\rm orb}$ (day)              &0.6169838471 (1)\\
$Q$ (day)                        &0.92(5) $\times$ 10$^{-10}$\\
$dP/dt$ (day yr$^{-1}$)          &1.09(8) $\times$ 10$^{-7}$\\
$dM/dt$ ($M_{\odot}$ yr$^{-1}$)  &0.40(5) $\times$ 10$^{-7}$\\
   \bottomrule
    \end{tabularx}
        \label{tab:o-c_parameters}
\end{table}%

\section{Simultaneous Analysis of Light and Radial Velocity Curves}
\label{LC}

The {\it TESS} data provides the main source of photometric data used in this study. Although the {\it Hipparcos}  satellite \citep{ESA1997} obtained the first light variation for the IS CMa system, {\it TESS} data have proven to be more precise. To present the system geometry in the most accurate way and with reliable parameters, the {\it TESS} light curve has been analysed simultaneously by taking the radial velocity data obtained by \citet{Ozkardes2010}. The effective temperature of the primary component was assumed at 7275 K given in the {\it Gaia} DR3 database \citep{Gaia2023}. After determining the effective temperature parameter, the {\it TESS} light curve was phased with the light elements which was also listed in Table \ref{tab:o-c_parameters} obtained in this study. The {\it TESS} light curve was analysed simultaneously with radial velocities using the Py\textsc{WD2015} \citep{Ozdarcan2020}, which is a new GUI software formed from the 2015 version of the Wilson–Devinney code \citep[W-D;][]{Wilson1971, Wilson1979, Wilson1990, HammeWilson2007, Wilson2008, WilsonHamme2014}. Solution approaches have been utilized for IS CMa in mode-3 (contact-state), which is classified as a contact system in the literature. Under convective outer envelope assumption ($T_\textit{\rm eff}$ $<$ 7500 K), the effective temperature values of the system's components, gravity darkening coefficients, and bolometric albedos were taken from $g_1$ = $g_2$ = 0.32 \citep{Lucy67} and $A_1$ = $A_2$ = 0.5 \citep{Rucinski69}, respectively. The linear bolometric and monochromatic limb-darkening coefficients were calculated from the values of \citet{Claret2017}. From the assumption of a circular orbit, the value of $e=0$ was chosen as the value to be used. Additionally, it was assumed that the components rotated synchronously; hence, the values $F_{1}=F_{2}=1$ were assigned to them. These parameters were fixed throughout the iterations. 

\par 
The parameters adjusted during the iterations are the mass ratio of the components of the system $q~(=M{_2}/M{_1})$, the systemic velocity ($V_{\gamma}$), the semi-major axis of binary system ($a$), orbital inclination ($i$), surface temperature of the secondary component ($T_2$), phase shift ($\phi$), the dimensionless surface equipotential ($\Omega$) of star 1 (star 2 is equivalent to that of star 1 for contact binaries ($\Omega_{1}=\Omega_{2}$)), and the monochromatic luminosity of the primary component ($L_1$). We noticed that the light curves are asymmetric \citep[O'Connell effect,][]{Connell51}, and we put forward a starspot model to explain this phenomenon. Thus, solutions have been obtained with the hypothesis of two cold spot on the hot component (primary component) to model the asymmetry in the light curve of the system. During the algorithm is running, for each  parameter corrections are much smaller than standard deviations ($\sigma$) the algorithm is considered solved. 

% Figure 2
\begin{figure}[hb!]
\centering\includegraphics[width=1\linewidth]{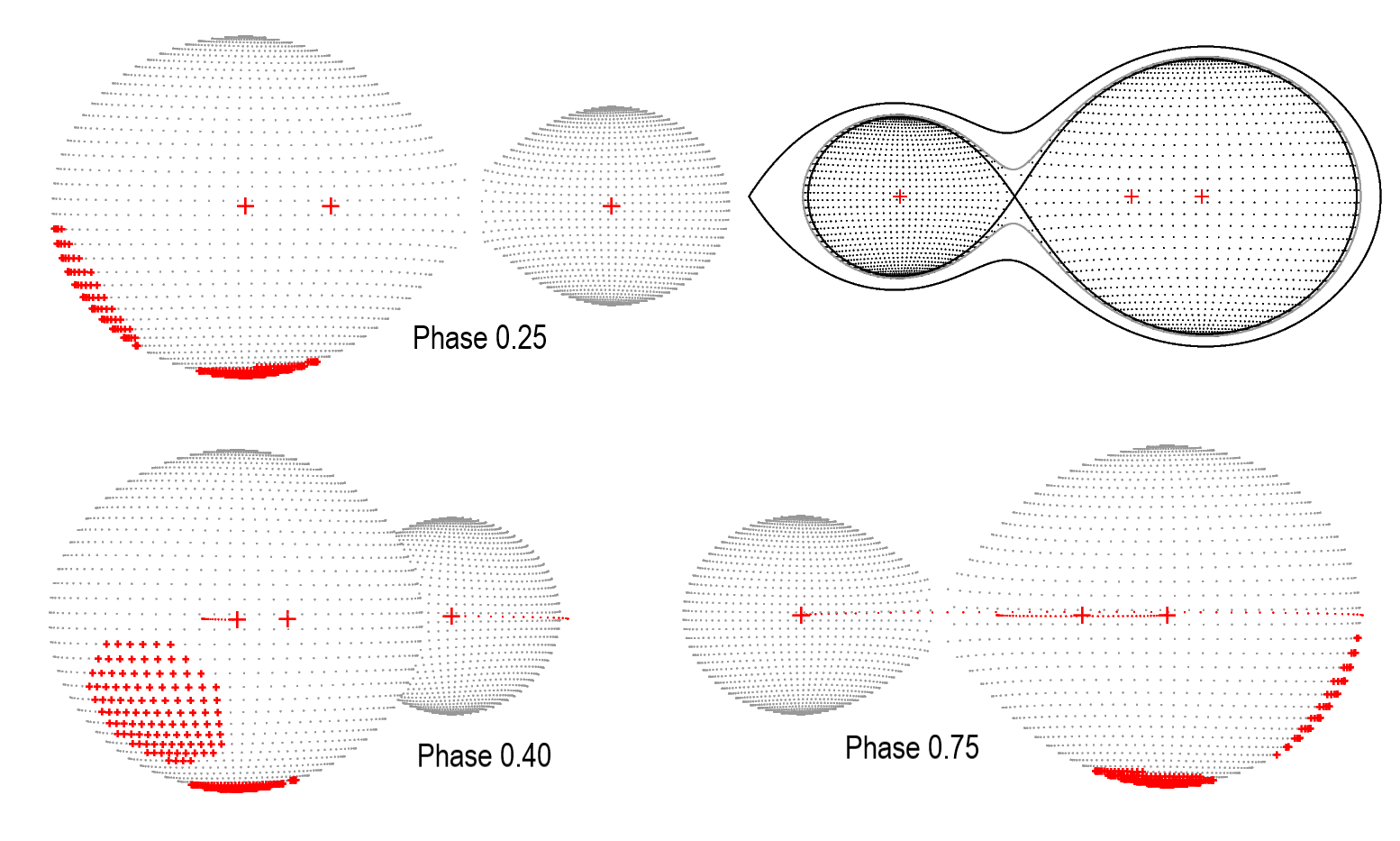}
\caption{The Roche geometry with spot configuration of IS CMa at different phases.}
    \label{fig:Roche}
\end{figure}

\begin{figure}[ht!]
\centering
\includegraphics[scale=0.33, angle=0]{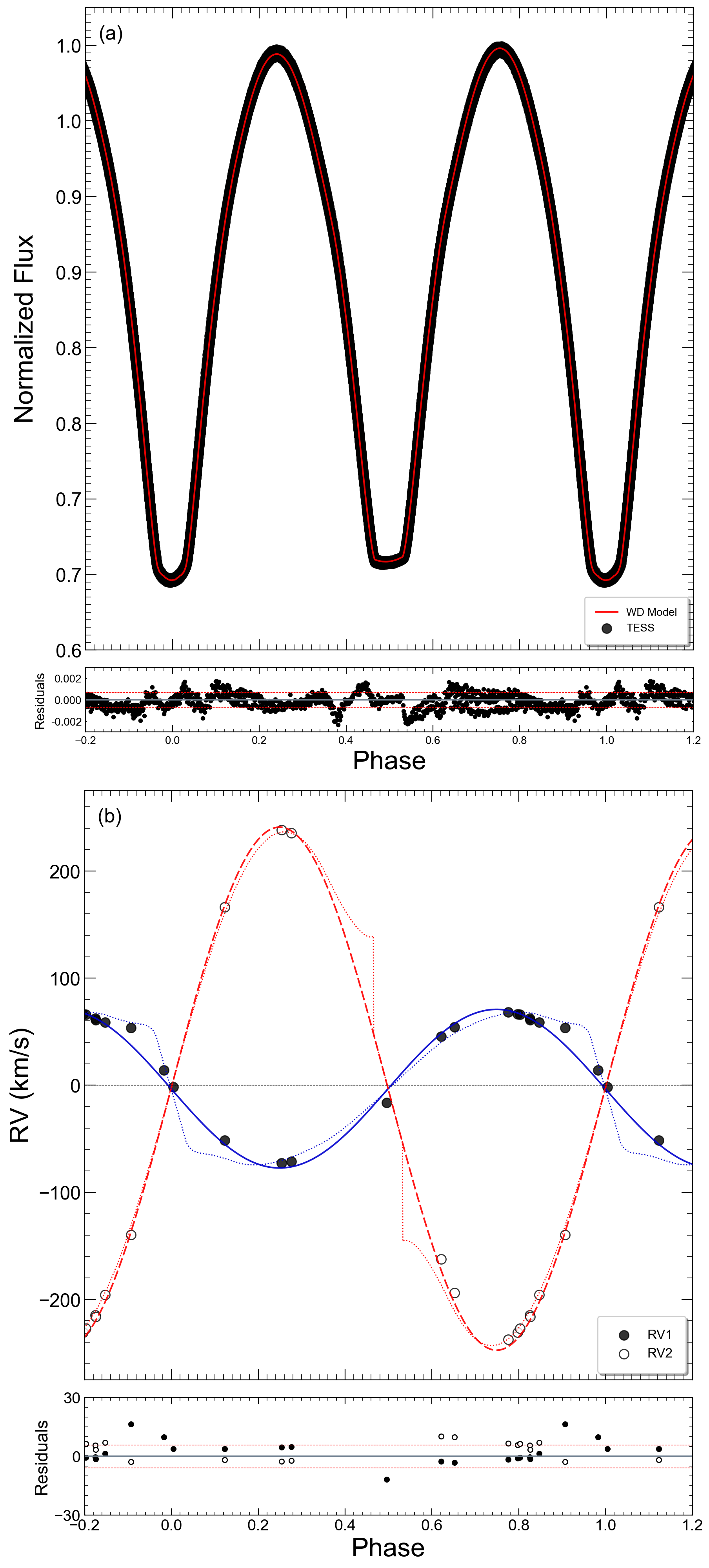}
\caption{(a) IS CMa {\it TESS} light and (b) radial curves with simultaneous solution of W-D theoretical curves. The model of the RVs includes both proximity effects (shown by the dotted lines) and a pure Keplerian solution (shown by the dashes and solid lines). The lower panels are the residuals from fits and red dots in these panels indicate $\pm 1\sigma$ levels.}
\label{fig:LC_RV}
\end{figure}

In this study, the mass ratio value and associated error were achieved with more precision using the simultaneous solution using radial velocities derived from \citet{Ozkardes2010} with {\it TESS} satellite data. Considering the error value of the mass ratio from the previous study and considering poor-quality data and distortions of the {\it Hipparcos} data \citep{McDonald2012}, it appears that the findings lack significance. The utilization of data from the {\it TESS} satellite-enabled us to acquire more accurate parameters for the system. In addition, a third light contribution has been determined to the total light of the {\it TESS} light curve, where is determined according to normalized phase at 0.25, $l_3=0.0912\pm 0.0010$. The third light contribution was notable due to the precision of the {\it TESS} space-based photometry with high sensitivity, which was initially achieved in this study \citep{Wang2020}. The spot position of the system is modelled with {\sc Binary Maker 3.0} Programme \citep{Binary02} and the Roche geometry given in Figure \ref{fig:Roche}. The fill-out factor for the IS CMa was determined using the equation $f$ = $(\Omega_{i}-\Omega)$/$(\Omega_{i}-\Omega_{\rm o})$ where $\Omega_{i}$ and $\Omega_{\rm o}$ are the inner and outer Lagrangian potentials, \citep{Kallrath2009}. The filling factor of the system was calculated as $f=20\%$. The calculated basic parameters are given by Table \ref{tab:basic} and the fundamental parameters are also presented in Table \ref{tab:salt}. Figure \ref{fig:LC_RV} presents the light and velocity curves with simultaneous W-D solutions. Additionally, we included radial velocities (RVs) with adjustments for proximity and eclipse corrections, as well as the purely Keplerian solution without proximity effects.

% Table 3
\begin{table}[b!]
\centering
\caption{Basic parameters of IS CMa.}
    \begin{tabularx}{\columnwidth} { 
   >{\raggedright\arraybackslash}X  >{\raggedleft\arraybackslash}X}
     \toprule
\text{Parameter}                         &\text{IS CMa}\\
\midrule
$P$ (day)                                & 0.6169838471$^1$\\
$T_{1}$ (K)                              & 7275 (fixed)$^2$\\
$T_{2}$ (K)                              & 6825 $\pm$ 201\\
Phase~Shift                              & -0.0009 $\pm$ 0.0001\\
$i~(^{\circ})$                           & 89.32 $\pm$ 0.03\\
$q~(=M{_2}/M{_1})$                       & 0.3032 $\pm$ 0.0002\\
$\Omega_{1} = \Omega_{2}$                & 2.4355 $\pm$ 0.0006\\
$a$~({\mbox{$R_{\odot}$}})               & 3.88 $\pm$ 0.02\\
$V_{\gamma}$  (km s$^{-1}$)              & -3.23 $\pm$ 1.56\\
$f (\%)$                                 & 20\\
$L_1/(L_1+L_2)_\textit{TESS}$            & 0.7821 $\pm$ 0.0002\\
$l_3$                                    & 0.0912 $\pm$ 0.0010\\
$r_1\textit {\rm (mean})$                & 0.498 $\pm$ 0.001\\
$r_2\textit {\rm (mean})$                & 0.293 $\pm$ 0.001\\
\hline
Co-Latitude$_{\rm Spot~1}$~$(^{\circ})$  & 117.89 $\pm$ 2.39\\
Longitude$_{\rm Spot~1}$~$(^{\circ})$    & 184.79 $\pm$ 0.16\\
Radius$_{\rm Spot~1}$~$(^{\circ})$       & 25 $\pm$ 2\\
Temperature~Factor$_{\rm Spot~1}$        & 0.89 $\pm$ 0.01\\
\hline
Co-Latitude$_{\rm Spot~2}$~$(^{\circ})$  & 174.80 $\pm$ 5.01\\
Longitude$_{\rm Spot~2}$~$(^{\circ})$    & 20.92 $\pm$ 0.10\\
Radius$_{\rm Spot~2}$~$(^{\circ})$       & 22 $\pm$ 1\\
Temperature~Factor$_{\rm Spot~2}$        & 0.90 $\pm$ 0.08\\
$\Sigma~\textit({O-C)}^2_i$              & 0.0003464355\\
\hline
    \end{tabularx}
\\
Ref: [1] This work, [2] \citet{Gaia2023}\\
\label{tab:basic}
\end{table}

It is revealed for the first time in this study that IS CMa is a system with having total eclipse. IS CMa is typically A type W UMa binary with mass ratio is $q=0.3032\pm0.0002$. A type W UMas are common for the massive component to have a hotter than the less massive one \citep{Binnendijk1970}. Analysing spectroscopic mass ratio (\textit{$q_{\rm sp}$}) with photometric one (\textit{$q_{\rm ph}$}), it has been noticed that there is a significant association between \textit{$q_{\rm sp}$} and \textit{$q_{\rm ph}$} for totally eclipsing overcontact binaries \citep{Senavci2016}. Nevertheless, it is true that IS CMa has a somewhat shallow fill out factor compared to other A type overcontacts. Additionally, it specifically exhibits two cold spots on the hot component. Studying in the literature suggests that the features of spots may also effect on the Roche lobes, so it is possible that impact on their fill out factor \citep{Yakut2005}.  

% Figure 3

% Table 4
\begin{table}[ht!]
\setlength{\tabcolsep}{2pt}
\renewcommand{\arraystretch}{1}
\centering
\caption{Fundamental parameters for IS CMa given in this study and in the literature}.
\begin{tabular}{lccc}
\hline
\text{Parameter}           &This Study & Literature & Ref.\\
%\midrule
\hline
$\alpha_{\rm J2000}$ (Sexagesimal)        & ~~06:20:56.41       &~~06:20:56.38      & [1] \\
$\delta_{\rm J2000}$ (Sexagesimal)        & -29:40:15.12        &-29:40:14.40       & [1] \\
$\mu_{\alpha}\cos \delta$ (mas yr$^{-1}$) & +40.735 $\pm$ 0.016 & +41.44 $\pm$ 0.65 & [1] \\
$\mu_{\delta}$ (mas yr$^{-1}$)            & -76.918 $\pm$ 0.017 &-77.97 $\pm$ 0.65  & [1] \\
$\varpi$ (mas)                            & 11.035 $\pm$ 0.017  & 10.01 $\pm$ 0.71  & [1] \\
$d$$\varpi$ (pc)                          & 90.62 $\pm$ 0.14    & 99.90 $\pm$ 7.09  & [1] \\
\hline
$M_1~(M_{\odot})$                         & 1.58 $\pm$ 0.01     & 1.68 $\pm$ 0.04   & [2] \\
$M_2~(M_{\odot})$                         & 0.48 $\pm$ 0.02     & 0.50 $\pm$ 0.02   & [2] \\
$R_1~(R_{\odot})$                         & 1.93 $\pm$ 0.01     & 2.00 $\pm$ 0.02   & [2] \\
$R_2~(R_{\odot})$                         & 1.14 $\pm$ 0.01     & 1.18 $\pm$ 0.03   & [2] \\
$T_1$ (K)                                 & 7275 (fixed)$^{[3]}$& 6800 $\pm$ 200    & [2] \\
$T_2$ (K)                                 & 6825 $\pm$ 201      & 6320 $\pm$ 270    & [2] \\
$\log L_1~(L_{\odot})$                    & 0.97 $\pm$ 0.01     & 0.88 $\pm$ 0.22   & [2] \\
$\log L_2~(L_{\odot})$                    & 0.40 $\pm$ 0.06     & 0.30 $\pm$ 0.10   & [2] \\
$\log g_1$ (cgs)                          & 4.06 $\pm$ 0.01     & 4.06 $\pm$ 0.03   & [2] \\
$\log g_2$ (cgs)                          & 4.01 $\pm$ 0.03     & 3.99 $\pm$ 0.02   & [2] \\
$M_{\rm Bol1}$ (mag)                      & 2.32 $\pm$ 0.13     & 2.54 $\pm$ 0.14   & [2] \\
$M_{\rm Bol2}$ (mag)                      & 3.75 $\pm$ 0.15     & 4.00 $\pm$ 0.21   & [2] \\
$M_{\rm V1}$ (mag)                        & 2.25 $\pm$ 0.13     & 2.59 $\pm$ 0.14   & [2] \\
$M_{\rm V2}$ (mag)                        & 3.67 $\pm$ 0.14     & 4.05 $\pm$ 0.21   & [2] \\
$V_1$ (mag)                               & 7.11                & ---               &     \\
$V_2$ (mag)                               & 8.53                & ---               &     \\
$A_{\rm d}$ (mag)                         & 0.024               & 0.034             & [2] \\
$E_{\rm d}(B-V)$ (mag)                    & 0.008               & 0.011             & [2] \\
$t$ (Gyr)                                 & 1.3 $\pm$ 0.1       & 1.59              & [2] \\
$a~(R_{\odot})$                           & 3.88 $\pm$ 0.02     & 3.95 $\pm$ 0.03   & [2] \\
$d$ (pc)                                  & 92.7 $\pm$ 6.5      & 87 $\pm$ 5        & [2] \\
$V_{\gamma}$ (km/s)                       &  -3.23 $\pm$ 1.56   & -2.75 $\pm$ 0.95  & [2] \\
${\rm [Fe/H]}$ (dex)                      & 0.01                & -0.36             & [4] \\
\hline
$U_{\rm LSR}$ (km/s)                      & 42.83 $\pm$ 0.80    & ---               &     \\
$V_{\rm LSR}$ (km/s)                      & -4.50 $\pm$ 1.24    & ---               &     \\
$W_{\rm LSR}$ (km/s)                      & 11.91 $\pm$ 0.51    & ---               &     \\
$S_{_{\rm LSR}}$ (km/s)                   & 44.68 $\pm$ 1.56    & ---               &     \\
$R_{\rm a}$ (pc)                          & 9184 $\pm$ 77       & ---               &     \\
$R_{\rm p}$ (pc)                          & 6717 $\pm$ 37       & ---               &     \\
$z_{\rm max}$ (pc)                        & 192 $\pm$ 10        & ---               &     \\
$e$                                       & 0.155 $\pm$ 0.002   & ---               &     \\
\hline
%\bottomrule
     \end{tabular}
    \newline
            Ref: [1] \citet{ESA1997}, [2] \citet{Ozkardes2010}, [3] \citet{Gaia2023}, [4] \citet{Nordstrom2004}
        \label{tab:salt}
\end{table}%

The dust map of \citet{Schlafly2011} was used to determine the $V$-band extinction value ($A_{\rm V}$) in the IS CMa direction. The Galactic coordinates ($l=237^{\rm o}.184639,~~b=-19^{\rm o}.166833$) of the system were determined as $A_{\infty}(b)=0.115\pm0.003$ mag in the direction of IS CMa using the dust extinction calculation tool on the NASA/IPAC website\footnote{https://irsa.ipac.caltech.edu/applications/DUST/}. The \citet{Bahcall1980} relation was used to calculate the reduced extinction ($A_{\rm d}(b)$) between the Sun and system:

\begin{equation}
A_{d}(b)=A_{\infty}(b)\Biggl[1-\exp\Biggl(\frac{-\mid
d\times \sin(b)\mid}{H}\Biggr)\Biggr].
\label{eq:indirgeme}
\end{equation}
where $b$ and $d$ are the Galactic latitude and distance of the star, respectively. $H$ is the scaleheight for the interstellar
dust which is adopted as 125 pc \citep{Marshall2006} and
$A_{\infty}(b)$ and $A_{\rm d}(b)$ are the total absorption for the dust map and the distance to the IS CMa, respectively \citep[see also,][]{Bilir2008a, Bilir2008b, Eker2009}. Moreover, the colour excess ($E_{\rm d}(B-V)$) to the direction of the system was also calculated using the $E_{\rm d}(B-V)=A_{\rm d}(b)~/~3.1$ relation. By substituting the distance of the system ($90.62\pm0.14$ pc) calculated from the trigonometric parallax ($\varpi=11.035\pm0.017$ mas) from \citet{Gaia2023} into Equation \ref{eq:indirgeme}, the colour excess and $V$-band extinction values were determined as $E_{\rm d}(B-V)=0.008$ and $A_{\rm d}=0.024$ mag, respectively. The parameters calculated in this section are listed in Table \ref{tab:salt}.

\section{Fundamental Parameters and Evolutionary Status}
The determination of the fundamental parameters of the component stars within the IS CMa system involved a simultaneous analysis of both light and radial velocity curves. In calculating the absolute parameters for IS CMa, the effective temperature $T_{\rm eff} = 5771.8\pm 0.7$ K, surface gravity $g = 27423.2\pm 7.9$ cm s$^{-2}$, and bolometric magnitude $M_{\rm Bol} = 4.7554\pm 0.0004$ mag as determined for the Sun by \citet{Pecaut2013} were used. The fundamental parameters and their uncertainties were obtained with the Py\textsc{WD2015} \citep{Ozdarcan2020}, and listed in detail in Table \ref{tab:salt}. The effective temperature of primary component of the system is assumed as $T_{1}=7275$ K taken from the {\it Gaia} DR3 database \citep{Gaia2023}. By analysing both photometric and spectroscopic data, the temperature of the second component was determined as $T_{2}=6825\pm 201$ K. Moreover, the masses and radii of the primary and secondary components were determined as $M_{1}= 1.58\pm 0.01\,M_\odot$, $M_{2}= 0.48\pm0.02\,M_\odot$, and $R_{1}=1.93\pm 0.01\,R_\odot$, $R_{2}= 1.14\pm 0.01\,R_\odot$, respectively. The luminosities of the component stars were calculated considering the bolometric correction ($BC$) coefficients of \citet{Eker2020}. The $BC$ values for primary and secondary components were taken as $BC_{1}=0.072$, $BC_{2}=0.080$ mag, respectively. The surface gravity parameters were determined for the primary and secondary components as log\,$g_{1} = 4.06 \pm 0.01$ and log\,$g_{2}=4.01 \pm 0.03$ in cgs, respectively.
 
\begin{figure}
\centering\includegraphics[width=1\linewidth]{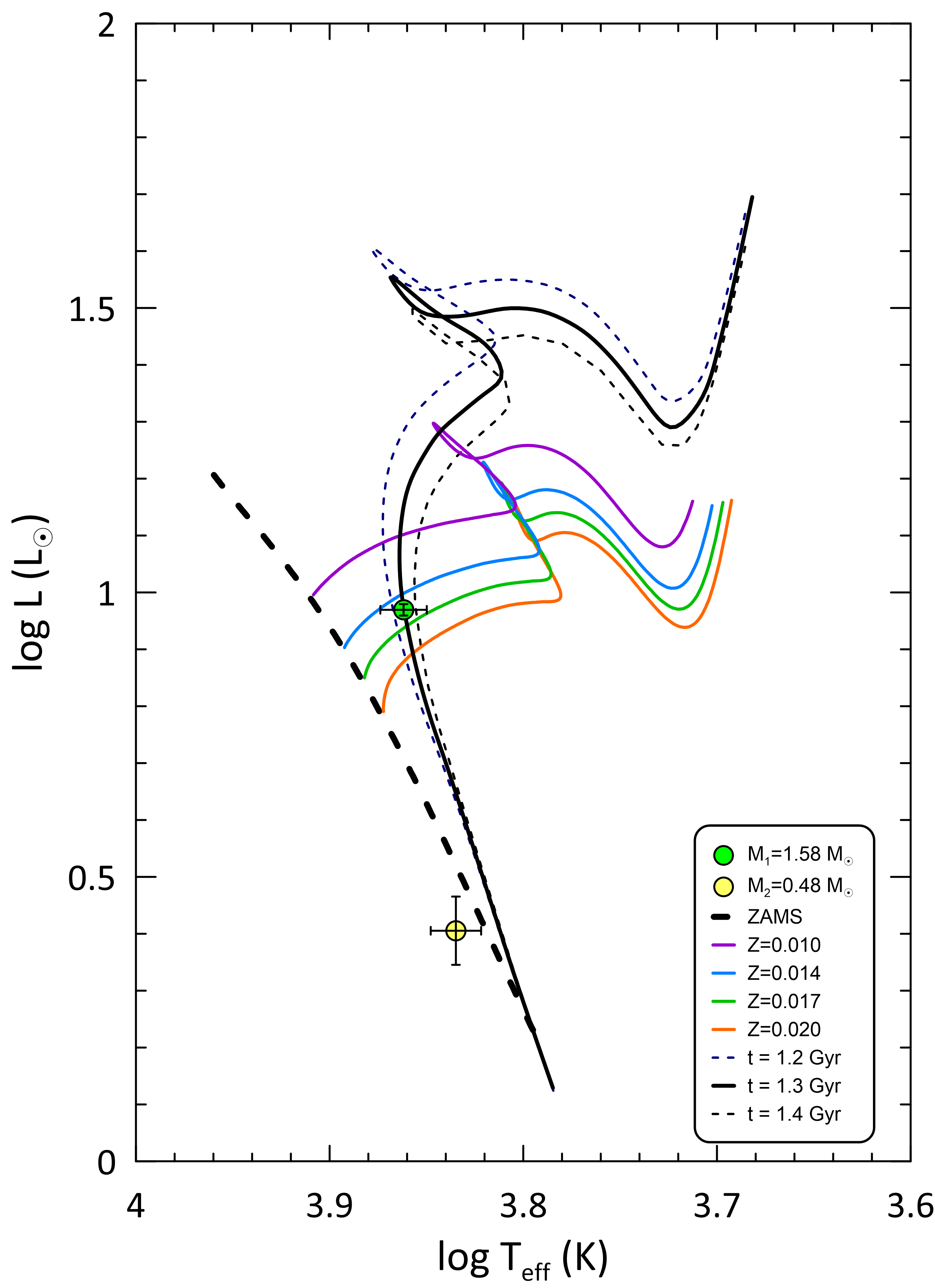}
\caption{Positions of the components of IS CMa on the HR diagram.}
\label{fig:HR-diagram}
\end{figure}

The accurate fundamental stellar parameters obtained provide a comprehensive insight into the binary component stars' evolutionary status and age. PARSEC isochrone \citep{Bressan2012} and mass tracks \citep{Chen2014, Tang2014} were used to investigate the evolutionary state of IS CMa system. The effective temperatures and luminosities of the components are plotted on a $\log L \times \log T_{\rm eff}$ diagram as shown in Figure \ref{fig:HR-diagram} along with the ZAMS curve. Since the metal abundances of the component stars in the IS CMa are not known, $M=1.60M_{\odot}$ PARSEC mass tracks with five different heavy element abundances ($Z=0.010, 0.014, 0.017$, and 0.020) were fitted to the primary component (Figure \ref{fig:HR-diagram}). Considering the position of the primary star in the HR diagram, it is located in $0.014 < Z < 0.017$ mass tracks. Therefore, the mean of these two abundances was taken for the determination of the heavy element abundance of the primary component and it was assumed as $Z=0.0155$. In this study, the heavy element abundance ($Z$= 0.0155) determined for IS CMa was converted to iron abundance ([Fe/H]= 0.01 dex) by Bovy's conversion formula\footnote{https://github.com/jobovy/isodist/blob/master/isodist/Isochrone.py} \citep[see more details,][]{Yontan2022, Dursun2024}. Although there is no metal abundance value calculated based on spectral analyses of IS CMa in the literature, \citet{Nordstrom2004} determined as a value [Fe/H]= -0.36 dex obtained from the Geneva-Copenhagen survey. The metal abundance given in the literature is found to be more metal poor than the one calculated in this study. However, it should be noted that the metal abundance given in the literature are valid for single stars.

The positions of the component stars in the $\log L \times \log T_{\rm eff}$ diagram indicate that the primary component of IS CMa lies in the main-sequence band, while the secondary component is more massive and extremely bright compared to a normal main-sequence star of the same mass. In addition, PARSEC mass tracks with $t=1.3$ Gyr were found to be the best fit for the primary component of the IS CMa. In this study, it was determined that the mass value of $1.58 M_{\odot}$ calculated for the primary component is compatible with the PARSEC mass track with a mass of 1.60 $M_{\odot}$. PARSEC isochrones were also fitted for the age determination of the system. The analyses showed that the $t=1.3\pm 0.1$ Gyr isochrones represent the age of the system quite accurately.

%Figure 4

\section{Position of IS CMa on the $\log J_{0}-\log M_{\rm tot}$ Diagram}
\begin{figure}[t!]
\centering\includegraphics[width=1\linewidth]{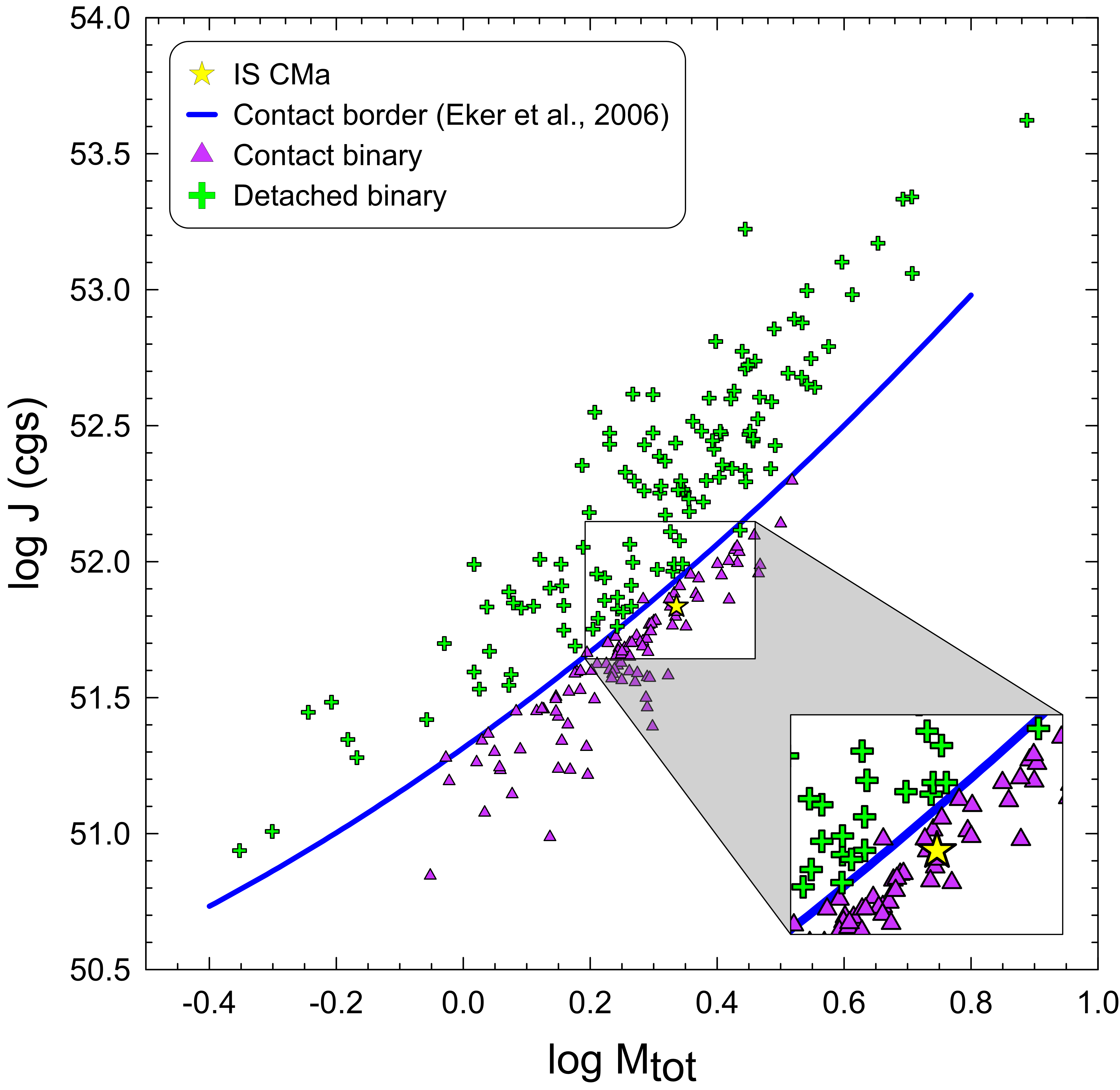}
\caption{Position of the IS CMa system in the $\log J\times \log M_{\rm tot}$ diagram. The blue curve shows the contact border obtained by \citet{Eker2006}. In addition, the plus, triangle, and star symbols denote detach systems, contact systems, and IS CMa. The zoomed-in insets in the figure represent the region of  condensation}.
\label{fig:contact-border}
\end{figure}

To investigate the contact situation of IS CMa, the orbital angular momentum of the system was calculated.  \citet{Eker2006} calculated the angular momentum and total masses of 119 chromospheric active stars and 102 W UMa type double stars and obtained the following relation by determining the separation of detached systems and contact systems on the $\log J \times \log M_{\rm tot}$ diagram with an empirical method:
\begin{eqnarray}
\log J_{\rm lim}=0.522(\log M_{\rm tot})^{2}+1.664(\log M_{\rm tot})+51.315,
\end{eqnarray}  
Here $J_0$ is the total angular momentum of a binary system and $M_{\rm tot}$ is the total mass of the components in the system. The following relation is used to calculate the angular momentum ($J_0$) of the system. 
%where $M$ is in solar units and $J_{lim}$ is in cgs.

\begin{eqnarray}
J_{0}={q\over (1+q)^2}\sqrt[3]{{G^2 \over 2\pi}M_{\rm tot}^5P},
\label{eq:J0}
\end{eqnarray} 
where $q$ is the mass ratio of the system, $G$ is the gravitational constant, $M_1$ and $M_2$ are the masses of the primary and secondary components, respectively, and $P$ is the orbital period of the system. The orbital angular momentum of IS CMa calculated from Equation \ref{eq:J0} is determined as $\log J_0 = 51.798$ cgs and the total logarithmic mass is determined as $\log M_{\rm tot} = 0.314$. The contact configuration of the system was analysed by plotting IS CMa on the $\log J_0 \times \log M_{\rm tot}$ diagram created by \citet{Eker2006}. As can be seen from Figure \ref{fig:contact-border}, IS CMa is located in a region of contact binaries. This supports the solution of IS CMa in the contact configuration (mode-3) and shows that it is a typical contact system.

% Figure 06

\section{Space Velocity Components and Galactic Orbital Parameters}
With the onset of the {\it Gaia} era, the astrometric data of the stars in the Solar neighborhood became more precisely identifiable, allowing the accurate calculation of the kinematic and dynamic orbital parameters of the stars \citep{Bilir2012, Koc2022, Doner2023}. To determine the space velocity components and Galactic orbital parameters of IS CMa analysed in this study, the proper-motion components ($\mu_{\alpha}\cos\delta,~\mu_{\delta}$) and trigonometric parallaxes ($\varpi$) of the system were taken from the {\it Gaia} DR3 database \citep{Gaia2023}, and the centre of mass velocities ($V_{\gamma}$) of IS CMa system were determined in this study are listed in Table \ref{tab:salt}. The space velocity components of the IS CMa were calculated using the {\sc galpy} code of \citet{Bovy2015} while the uncertainties of space velocity components were determined via the algorithm of \citet{Johnson1987}. The space velocity components contain some biases due to the positions of stars in the Milky Way and observations from the Sun. To eliminate these biases, the space-velocity components of the stars should be corrected by differential rotation and local standard rest (LSR) \citep{Karatas2005, Tuysuz2014, Coskunoglu2012, Bilir2016}. The equations of \citet{Mihalas1981} are used for the differential rotational corrections of IS CMa, and the velocity corrections for the ($d_{\rm U}, d_{\rm V}$) space-velocity components of the systems are calculated as (-1.96, -0.11) km s$^{-1}$. Since the $W$ space-velocity component is not affected by the differential rotation correction, no correction is applied to the $W$ space-velocity component of the IS CMa. For the LSR correction, the values of \citet{Coskunoglu2011} $(U, V, W)_{\odot}=(8.83\pm 0.24, 14.19\pm 0.34, 6.57\pm 0.21)$ kms$^{-1}$ were considered and the LSR values were extracted from the space-velocity components for which a differential velocity correction was applied. The relation $S_{\rm LSR}^2=U_{\rm LSR}^2+V_{\rm LSR}^2+W_{\rm LSR}^2$ was used to calculate the total space velocity ($S_{\rm LSR}$) of IS CMa and the results are given in Table \ref{tab:salt}. Considering the total space velocities and space velocity component values of the IS CMa, it is consistent with the value given for the young thin-disc population \citep{Leggett1992}.

% Figure 5
\begin{figure}[h!]
\centering\includegraphics[width=1\linewidth]{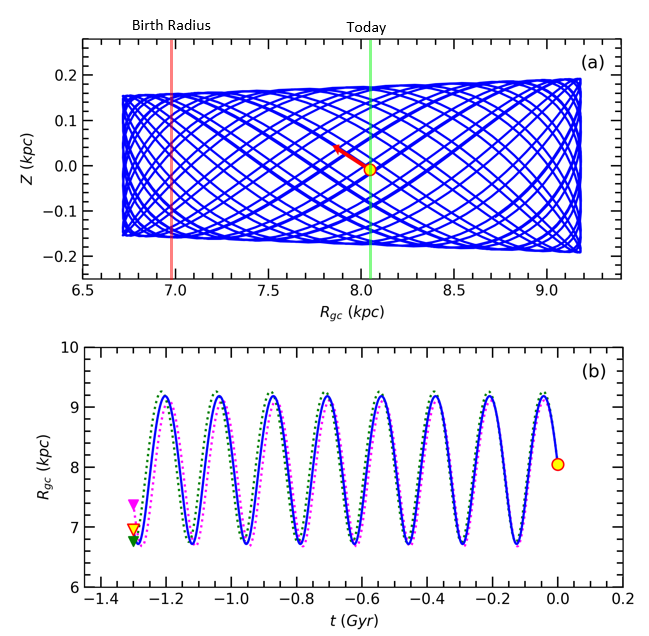}
\caption{The Galactic orbits and birth radii of IS CMa in the $Z \times R_{\rm gc}$ (a) and $R_{\rm gc} \times t$ (b) diagrams. The filled yellow circles and triangles show the today and birth positions, respectively. The red arrow is the motion vector of IS CMa in today. The green and pink dotted lines show the orbit when errors in input parameters are considered, while the green and pink filled triangles represent the birth locations of the IS CMa based on the lower and upper error estimates.}
\label{fig:galactic_orbits}
\end{figure} 

The {\sc galpy} code \citep{Bovy2015} was also used to calculate the Galactic orbital parameters of the IS CMa. {\sc MWPotential2014} was used for the Galactic potentials needed for the Galactic orbit calculations, which was created specifically for the Milky Way. For the systems to form closed orbits around the center of the Galaxy, a time scale of 3 Gyr in 5 Myr steps was investigated. As a result of the analyses of the orbital calculations, the perigalactic ($R_{\rm p}$), apogalactic ($R_{\rm a}$), maximum distance from the Galactic plane ($z_{\rm max}$), and eccentricity ($e$) of the Galactic orbits were determined, and are also listed in Table \ref{tab:salt}. The positions of the system according to their distances from the Galactic center ($R_{\rm gc}$) and perpendicular to the Galactic plane ($z$) at different times are shown in Figure \ref{fig:galactic_orbits}a. The {\sc galpy} results show that the IS CMa has a slightly flattened Galactic orbit. Moreover, the fact that the system is at $z=-30$ pc ($z=d\sin(b)$) from the Galactic plane is an indication that IS CMa may belong to the young thin-disc component of Milk Way \citep{Guctekin2019}.

The Galactic orbits for the IS CMa system on the $z \times R_{\rm gc}$ and $R_{\rm gc} \times t$ planes are represented in Figure \ref{fig:galactic_orbits}. The panels in Figure \ref{fig:galactic_orbits} show side views of the IS CMa motions as functions of distance from the Galactic center and the Galactic plane, respectively \citep{Tasdemir2023, Yontan2023b}. In Figure \ref{fig:galactic_orbits}b the birth and present-day positions for the IS CMa are shown with yellow-filled triangle and circle, respectively \citep{Yontan2023a}. The eccentricity of the orbit of the IS CMa does not exceed the value of 0.16. The distances from the Galactic plane reach out maximum at $z_{\rm max}=192\pm10$ pc for IS CMa. These findings show that the binary system belongs to the young thin disc of the Milky Way. The birthplace of the IS CMa was also investigated by running the binary system age ($t=1.3\pm0.1$ Gyr) calculated in this study back in time using the {\sc galpy} program. The birth radius of the binary system was determined as $R_{\rm Birth}=6.95\pm0.31$ kpc. These results show that the IS CMa was born in the metal-rich region inside the Solar circle.

\section{Discussion and conclusions}

In this study, ground-based spectroscopic and space-based precision photometric observations of IS CMa were combined to determine the orbital period variation of the system, the fundamental astrophysical parameters of the component stars, and the kinematic and dynamical orbital parameters. The absolute parameters of the IS CMa components were calculated with their fundamental astrophysical relations using parameters obtained from simultaneous analysis of the light and RV curves. The temperature difference between the components is calculated to be about 450 K, which suggests that the system may have almost reached thermal equilibrium. Furthermore, the degree of contact of IS CMa was calculated to be $f=20\%$. These results indicate that the system is a type A W UMa. The mass ratio of the system was calculated as $q = 0.303$, which is in good agreement with the $q=0.297$ value given by \citet{Ozkardes2010}, who studied IS CMa with ground-based photometric and spectroscopic analyses. In addition, considering the temperatures of the components, the $BC$ values of the components were taken from the table given by \citet{Eker2020}. The masses and radii of the primary and secondary components for IS CMa were determined as $M_{1}= 1.58\pm 0.01\,M_\odot$, $M_{2}= 0.48\pm0.02\,M_\odot$, and $R_{1}=1.93\pm 0.01\,R_\odot$, $R_{2}= 1.14\pm 0.01\,R_\odot$, respectively. Moreover, the photometric distance of the system was calculated as $92.7\pm6.5$ pc, and this value is consistent with the value from {\it Gaia} DR3 database \citep{Gaia2023}. In this study, the precise trigonometric parallax measurement of IS CMa given in the {\it Gaia} DR3 catalogue and the $V$-band apparent magnitudes were used to test the temperatures of the component stars (see Table \ref{tab:salt}). The absolute magnitudes of each component were determined using the distance modul relation between the apparent magnitude of the component stars and the {\it Gaia} distance, and the $BC$ coefficients with $M_{\rm Bol}$ values corresponding to these absolute magnitudes were calculated. Then, the luminosities of the component stars were calculated by putting the calculated $M_{\rm Bol}$ of stars and $M_{\rm Bol, \odot}$ of the Sun \citep[][]{Pecaut2013} into the Pogson relation. Since the radii of the primary and secondary stars in the system are also known, via $L=4\pi R^2\sigma T^4$ relation, their temperatures were determined as 7442 and 6993 K, respectively. The differences between the temperatures determined based on {\it Gaia} distances and those calculated in this study are about $\Delta T_{1,2} = 170$ K. The results are consistent within the temperature uncertainties in Table \ref{tab:salt}.

The stellar parameters obtained in this study are almost in agreement with the results of \citet{Ozkardes2010}, who determined the fundamental parameters of IS CMa by combining {\it Hipparcos} photometric and spectroscopic observations. However, in general, the masses and radii of the component stars calculated in this paper are smaller than the values given by \citet{Ozkardes2010}. Although the calculated distances are consistent within the uncertainties, there is a difference of approximately 6.2\%. While this study used precise {\it TESS} photometric data, \citet{Ozkardes2010} used 141 photometric data taken from {\it Hipparcos} satellite \citep{ESA1997} is considerably less than this work. These may be the reasons for the differences in the analysis results.

The orbital period variations of the contact binary system IS CMa were examined for the first time in this work. All times have been searched in the literature, and new minimum times have been derived using {\it TESS} data. We modelled the IS CMa system focusing on the hot component, assuming the presence of two cold spots to account for asymmetries in the light curve. Additionally, a third light contribution was identified, which was not previously determined due to the satellite data's sensitivity. Given the dynamic effects, sinusoidal variations should be expected in addition to the parabolic variation of the system when observing changes in the orbital period. Changes caused by a certain magnetic cycle or a third body should result in observable periodic distortions in the {\it O-C} curves. More observation minima times are required to elucidate the modulations that may exist in the IS CMa system. Based on the limited data available, we can conclude that the system's orbital period has increased. Although the current orbital period of IS CMa appears to be increasing, the dynamical structure of the system demonstrates that there may also be a long-term change in addition to the parabolic behaviour. This can be further investigated through observations in the coming decades \citep{Qian2001, Gazeas2008, Liu2018}.

Detailed kinematic and dynamical orbital analyses of the system were performed for the first time in this study. Kinematic analyses show that IS CMa belongs to the young thin-disc population. In addition, $z_{\rm max} = 192\pm10$ pc and orbital eccentricity $e=0.155\pm0.002$ calculated as a result of dynamic orbital analyses also support the classification of the population of the binary system. In this study, the birth radius of the system $R_{\rm Birth}= 6.95\pm0.31$ kpc was calculated considering the age value calculated for IS CMa. This finding indicates that IS CMa formed in a region inside the Solar neighbourhood \citep{Banks2020, Yontan2023c}.

\section*{Acknowledgments}
We thank the anonymous referee for his/her insightful and constructive suggestions, which significantly improved the paper. This study has been supported in part by the \fundingAgency{Scientific and Technological Research Council (TÜBİTAK)} \fundingNumber{119F072}. This study is a part of the Ph.D. theses of Serkan Evcil. “Funding for the TESS mission is provided by NASA’s Science Mission directorate. This work has made use of data from the European Space Agency (ESA) mission \emph{Gaia}\footnote{https://www.cosmos.esa.int/gaia}, processed by the \emph{Gaia} Data Processing and Analysis Consortium (DPAC)\footnote{https://www.cosmos.esa.int/web/gaia/dpac/consortium}. Funding for DPAC has been provided by national institutions, in particular, the institutions participating in the \emph{Gaia} Multilateral Agreement. This research has made use of NASA’s Astrophysics Data System and the SIMBAD database, operated at CDS, Strasbourg, France. The MAST, Atlas O-C and O-C gateway databases were used in this paper. Thus, we express our gratitude to the working groups of MAST, Atlas O-C and O-C gateway.

\subsection*{Author contributions}

\textbf{Conception/Design of study}: SE, SB;\\ 
\textbf{Data Acquisition}: SE, RC; \\
\textbf{Data Analysis/Interpretation}: SA, SE; \\
\textbf{Drafting Manuscript}: SE, NA, SB, SA, RC; \\
\textbf{Critical Revision of Manuscript}: SE, NA, SB, SA, RC; \\
\textbf{Final Approval and Accountability}: SE, NA, SB, SA, RC.

\subsection*{Financial disclosure}

None reported.

\subsection*{Conflict of interest}

The authors declare no potential conflict of interests.

\nocite{*}% Show all bib entries - both cited and uncited; comment this line to view only cited bib entries;
%\bibliography{Wiley-ASNA}%

%\section*{Author Biography}
%(if applicable)
%\begin{biography}{\includegraphics[width=60pt,height=70pt,draft]{empty}}{\textbf{Author Name.} This is sample author biography text this is sample author biography text this is sample author biography text this is sample author biography text this is sample author biography text this is sample author biography text this is sample author biography text this is sample author biography text this is sample author biography text .}
%\end{biography}

\end{document}